\documentclass[aps,prd,preprintnumbers,groupedaddress,nofootinbib,amssymb,notitlepage,eqsecnum]{revtex4-1}
\usepackage{here}
\usepackage[dvipdfmx]{graphicx}
\usepackage{amsmath,amsthm,amssymb}
\usepackage{bm}
\usepackage{color}
\usepackage[dvipsnames]{xcolor}

\usepackage{amsfonts}
\usepackage{dcolumn}
\usepackage{hyperref}
\allowdisplaybreaks[1]
\usepackage{stackengine}

\newcommand{\be}{\begin{equation}}  
\newcommand{\ee}{\end{equation}}
\newcommand{\ba}{\begin{eqnarray}}
\newcommand{\ea}{\end{eqnarray}}

\newcommand{\rd}{{\rm d}}

\newcommand{\tp}{\dot{\phi}}
\newcommand{\cH}{{\cal H}}

\newcommand{\bem}{\begin{bmatrix}}
\newcommand{\eem}{\end{bmatrix}}
\newcommand{\Mpl}{M_{\rm Pl}}

\def\baoonly{0}

\allowdisplaybreaks

\begin{document}

\preprint{WUAP-23-10}

\title{Observational constraints on interactions between 
dark energy and dark matter\\
with momentum and energy transfers}

\author{Xiaolin Liu$^{1,5,6}$, Shinji Tsujikawa$^{2}$, Kiyotomo Ichiki$^{3,4,5}$}

\affiliation{
$^{1}$Department of Astronomy, Beijing Normal University,
Beijing 100875, China, \\
$^{2}$Department of Physics, Waseda University, 3-4-1 Okubo, Shinjuku, Tokyo 169-8555, Japan, \\
$^{3}$Kobayashi-Maskawa Institute for the Origin of Particles and the Universe, Nagoya University, Furocho, Chikusa-ku, Nagoya, Aichi 464-8602, Japan, \\
$^{4}$Institute for Advanced Research, Nagoya University, Furocho, Chikusa-ku, Nagoya, Aichi 464-8602, Japan, \\
$^{5}$Graduate School of Science, Division of Particle and Astrophysical Science, Nagoya University, Furocho, Chikusa-ku, Nagoya, Aichi 464-8602, Japan, \\
$^{6}$Institudo de Física Téorica UAM-CSIC, Universidad Autónoma de Madrid, 28049, Spain
}

\begin{abstract}

We place observational constraints on a dark energy (DE) model 
in which a quintessence scalar field $\phi$ is coupled to 
dark matter (DM) through momentum and energy exchanges.
The momentum transfer is weighed by an interaction between the 
field derivative and DM four velocity with a coupling 
constant $\beta$, whereas the energy exchange is characterized 
by an exponential scalar-field coupling to the DM density 
with a coupling constant $Q$. 
A positive coupling $\beta$ leads to the suppression for 
the growth of DM density perturbations at low redshifts, 
whose property offers a possibility for resolving
the $\sigma_8$ tension problem.
A negative coupling $Q$ gives rise to a $\phi$-matter-dominated 
epoch, whose presence can reduce the sound horizon around the 
Cosmic Microwave Background (CMB) decoupling epoch.  
Using the data of Planck 2018, 12-th Sloan Digital Sky Survey, 
Phantheon supernovae samples, and 1-year dark energy survey, 
we find that the two couplings are constrained to be 
$\beta=0.332^{+1.246}_{-0.237}$ and 
$Q =-0.0312^{+0.0312}_{-0.0085}$ at 68\,\% 
confidence level (CL). Thus, there is 
an interesting observational signature of 
the momentum exchange ($\beta \neq 0$) 
between DE and DM, with a peak of the 
probability distribution of the energy transfer 
coupling at $Q<0$. 

\end{abstract}

\date{\today}


\maketitle

\section{Introduction}
\label{introsec}

Revealing the origin of the dark sector in our Universe is 
an important challenge for the modern 
cosmology \cite{Jungman:1995df,Bertone:2004pz,Peebles:2002gy,Copeland:2006wr,Silvestri:2009hh,Kase:2018aps,Ezquiaga:2018btd}. 
Dark energy (DE) accelerates the current Universe, 
while cold dark matter (CDM) is the main source for the 
formation of large-scale structures. 
The origin of DE can be a cosmological constant 
$\Lambda$ \cite{Weinberg:1988cp,Sahni:1999gb,Padmanabhan:2002ji,Carroll:2000fy}, 
but it is theoretically challenging to naturally 
explain its small value from the vacuum energy 
arising from particle physics \cite{Martin:2012bt,Padilla:2015aaa}.
Instead, there have been many attempts for constructing 
DE models with dynamical propagating degrees of freedom 
such as scalar fields, vector fields, and massive 
gravitons (see Refs.~\cite{DeFelice:2010aj,Clifton:2011jh,Joyce:2014kja,Koyama:2015vza,Ishak:2018his,Heisenberg:2018vsk} for reviews). 
Among them, the scalar-field DE, which is 
dubbed quintessence \cite{Fujii:1982ms,Ford:1987de,Ratra:1987rm,Wetterich:1987fm,Chiba:1997ej,Ferreira:1997au,Caldwell:1997ii,Zlatev:1998tr}, 
is one of the simplest models 
which can be distinguished from the cosmological 
constant through its time-varying 
equation of state (EOS) $w_{\rm DE}$.

From the observational side, we have not yet found compelling 
evidence that quintessence is favored over the cosmological 
constant. In particular, the joint analysis based on the data 
of supernovae Ia (SN Ia), baryon acoustic oscillations (BAO), 
and the cosmic microwave background (CMB) showed that the 
quintessence EOS needs to be close to $-1$ at low redshifts \cite{Chiba:2012cb,Tsujikawa:2013fta,Durrive:2018quo,Planck:2018vyg,Akrami:2018ylq}.
Hence it is difficult to distinguish between quintessence 
and $\Lambda$ from the information of $w_{\rm DE}$ alone.
At the level of perturbations, the $\Lambda$CDM model 
has a so-called $\sigma_8$ tension for the amplitude of matter density 
contrast between the Planck CMB data \cite{Planck:2018vyg} 
and low-redshift probes like shear-lensing \cite{Heymans:2012gg,Hildebrandt:2016iqg,Abbott:2017wau} and 
redshift-space distortions \cite{Samushia:2013yga,Macaulay:2013swa}. 
For both $\Lambda$ and quintessence, the effective gravitational coupling $G_{\rm eff}$ on scales relevant to the growth of 
large-scale structures is equivalent to the Newton constant $G$. 
Then, the problem of the $\sigma_8$ tension cannot be addressed 
by quintessence either. 
Moreover, for both $\Lambda$ and quintessence, 
there is the tension of today's Hubble expansion rate $H_0$ 
between the CMB data and low-redshift measurements \cite{Aghanim:2018eyx,Riess:2018uxu,Wong:2019kwg,Riess:2020fzl,Verde:2019ivm,DiValentino:2021izs,Perivolaropoulos:2021jda,Freedman:2021ahq}.

If we allow for a possibility of interactions between DE 
and DM, the cosmic expansion and growth histories can be 
modified in comparison to the $\Lambda$CDM model. 
One example of such couplings corresponds to an energy 
exchange between DE and DM through an interacting 
Lagrangian $L_{\rm E}=-(e^{Q \phi/\Mpl}-1) 
\rho_c$ \cite{Wetterich:1994bg,Amendola:1999er,Frusciante:2018tvu,Kase:2019veo}, 
where $Q$ is a coupling constant, $\Mpl$ is the reduced 
Planck mass, and $\rho_c$ is the CDM density.
The similar type of couplings arises from Brans-Dicke 
theories \cite{Brans:1961sx} 
after transforming the Jordan-frame action to that 
in the Einstein frame \cite{Amendola:1999qq,Khoury:2003rn,Tsujikawa:2008uc}.
In the presence of such an energy transfer, it is possible 
to realize a so-called $\phi$-matter-dominated 
epoch ($\phi$MDE) \cite{Amendola:1999er} 
in which the DE (scalar field) density parameter takes 
a nonvanishing constant value $\Omega_{\rm DE}=2Q^2/3$. 
The presence of the $\phi$MDE can reduce the sound horizon 
at CMB decoupling \cite{Pettorino:2013oxa,Planck:2015bue,Gomez-Valent:2020mqn}, which may offer a possibility for alleviating 
the $H_0$ tension. 
On the other hand, the effective gravitational 
coupling of CDM is given by 
$G_{\rm eff}=G(1+2Q^2)$ \cite{Amendola:2003wa,Tsujikawa:2007gd}, 
which is larger than $G$.
This property is not welcome for reducing the $\sigma_8$ 
tension, as we require that $G_{\rm eff}<G$ 
to address this problem.  

The scalar field can also mediate the momentum exchange 
with CDM through a scalar product
$Z=u_c^{\mu} \nabla_\mu \phi$ \cite{Pourtsidou:2013nha,Boehmer:2015sha,Skordis:2015yra,Koivisto:2015qua,Pourtsidou:2016ico,Dutta:2017kch,Linton:2017ged,Kase:2019veo,Kase:2019mox,Chamings:2019kcl,Amendola:2020ldb,Kase:2020hst,Linton:2021cgd}, 
where $u_c^{\mu}$ is a CDM four velocity and 
$\nabla_\mu \phi$ is a covariant derivative of $\phi$.
If we consider an interacting Lagrangian of the form 
$L_{\rm M}=\beta Z^2$, where $\beta$ is a coupling 
constant, the modification to the background equations 
arises only through a change of the kinetic term 
$\dot{\phi}^2/2 \to (1+2\beta)\dot{\phi}^2/2$ 
in the density and pressure of 
$\phi$ \cite{Pourtsidou:2013nha,Pourtsidou:2016ico}.
At the level of perturbations, the Euler equation is 
modified by the momentum transfer, while the continuity 
equation is not affected.
For $\beta>0$, the conditions for the absence of 
ghosts and Laplacian instabilities of scalar and 
tensor perturbations are consistently 
satisfied \cite{Kase:2019mox}.
In this case, the effective gravitational coupling 
of CDM is smaller than $G$ at low redshifts \cite{Pourtsidou:2013nha,Pourtsidou:2016ico,Kase:2019veo,Kase:2019mox}.
Then, there is an intriguing possibility for reducing 
the $\sigma_8$ tension by the momentum 
transfer \cite{Pourtsidou:2016ico,Linton:2017ged,Chamings:2019kcl,Linton:2021cgd}.

An interacting model of DE and DM with both momentum 
and energy transfers was proposed in Ref.~\cite{Amendola:2020ldb} 
as a possible solution to the problems 
of $\sigma_8$ and $H_0$ tensions. 
This is described by the interacting Lagrangian 
$L_{\rm int}=\beta Z^2-(e^{Q \phi/\Mpl}-1)\rho_c$ 
with a canonical scalar field $\phi$ having a potential 
$V(\phi)$. Since the model has an explicit Lagrangian, 
the perturbation equations of motion are unambiguously 
fixed by varying the corresponding action with 
respect to the perturbed variables. 
We would like to stress that this is not the case 
for many interacting DE and DM models in which  
the background equations alone are modified by 
introducing phenomenological couplings \cite{Dalal:2001dt,Zimdahl:2001ar,Chimento:2003iea,Wang:2005jx,Amendola:2006dg,Guo:2007zk,Valiviita:2008iv,Salvatelli:2014zta,Kumar:2016zpg,DiValentino:2017iww,Yang:2018euj,Pan:2019gop,DiValentino:2019ffd,DiValentino:2019jae,
Salzano:2021zxk,Poulin:2022sgp}. 
We note however that there are some other models with 
concrete Lagrangians or energy-momentum tensors based on 
interacting fluids of DE and DM \cite{Asghari:2019qld,Jimenez:2020ysu,BeltranJimenez:2020qdu,BeltranJimenez:2021wbq} or on vector-tensor theories \cite{DeFelice:2020icf}.

In Ref.~\cite{Amendola:2020ldb}, it was anticipated that 
the momentum transfer associated with the coupling $\beta$ 
may address the $\sigma_8$ tension due to the suppression of growth 
of matter perturbations and that the energy transfer 
characterized by the coupling $Q$ may ease the $H_0$ 
tension by the presence of the $\phi$MDE.
While the gravitational attraction is enhanced 
by the energy transfer, the decrease  
of $G_{\rm eff}$ induced by the coupling $\beta$ can 
overwhelm the increase of $G_{\rm eff}$ induced 
by the coupling $Q$ \cite{Amendola:2020ldb,Kase:2020hst}.
We also note that the coupling $\beta$ does not remove 
the existence of the $\phi$MDE at the background level. 
These facts already imply that nonvanishing values of 
couplings may be favored, but we require 
a statistical analysis with actual observational data to 
see the signatures of those couplings.

In this paper, we perform the Markov chain Monte Carlo 
(MCMC) analysis of the interacting model of DE and DM 
with momentum and energy transfers mentioned above. 
For this purpose, we exploit the recent data of 
Planck CMB \cite{planck_2018}, 
12-th Sloan Digital Sky Survey (SDSS) \cite{sdss_dr12}, 
Phantheon supernovae samples \cite{pantheon_2018}, 
and 1-year dark energy survey (DES) \cite{DES_1yr}.
We show that the nonvanishing value of $\beta$ is 
statistically favoured over the case $\beta=0$, 
so there is an interesting signature of the momentum 
transfer between DE and DM. 
For the energy transfer, the probability distribution 
of the coupling has a peak at $Q<0$. 
The $Q=0$ case is also consistent with the data 
at 68\,\% CL, so the signature of energy transfer is 
not so significant compared to that of momentum transfer.
Today's Hubble constant is constrained to be 
$H_0=68.20^{+0.54}_{-0.55}$ km/s/Mpc (68\,\% CL), 
which is not much different from the bound 
derived for the $\Lambda$CDM model with the above 
data sets. Like most of the models proposed in the literature, 
our coupled DE-DM scenario does not completely resolve 
the Hubble tension problem present in the current 
observational data.

This paper is organized as follows. 
In Sec.~\ref{backsec}, we revisit the background 
dynamics in our interacting model of DE and DM.
In Sec.~\ref{persec}, we show the full linear 
perturbation equations of motion and discuss 
the stability and the effective gravitational 
couplings of nonrelativistic matter. 
In Sec.~\ref{implementation}, we explain the 
methodology of how to implement the background 
and perturbation equations in the \texttt{CAMB} code. 
We also discuss the impact of our model 
on several observables.
In Sec.~\ref{results}, we present our MCMC results 
and interpret constraints on the model parameters.
Sec.~\ref{consec} is devoted to conclusions.
Throughout the paper, we work in the natural unit system, 
i.e., $c=\hbar=k_B=1$.

\section{Background equations of motion}
\label{backsec}

We consider a DE scalar field $\phi$ interacting with 
CDM through energy and momentum transfers. 
We assume that $\phi$ is a canonical field with the kinetic 
term $X=-(1/2)\nabla^{\mu}\phi \nabla_{\mu}\phi$ 
and the exponential potential $V(\phi)=V_0 e^{-\lambda \phi/\Mpl}$, 
where $V_0$ and $\lambda$ are constants. 
The choice of the exponential potential is not essential 
for the purpose of probing the DE-DM couplings, but we can 
choose other quintessence potentials like the 
inverse power-law type 
$V(\phi)=V_0 \phi^{-p}$ \cite{Pettorino:2013oxa,Planck:2015bue,Gomez-Valent:2020mqn}.
The energy transfer is described by the interacting Lagrangian 
$L_{\rm E}=-(e^{Q \phi/\Mpl}-1)\rho_c$, 
where $Q$ is a coupling constant and $\rho_c$ is the 
CDM density. In the limit that $Q \to 0$, we have $L_{\rm E} \to 0$.
The momentum transfer is weighed by the interacting Lagrangian 
$L_{\rm M}=\beta Z^2$, where $\beta$ is a coupling constant 
and $Z$ is defined by 
\be
Z=u_c^{\mu} \nabla_{\mu}\phi\,,
\ee
where $u_c^{\mu}$ is the CDM four velocity. 
For the gravity sector, we consider Einstein gravity 
described by the Lagrangian of a Ricci scalar $R$. 
Then, the total action is given by \cite{Amendola:2020ldb}
\be
{\cal S}=\int {\rm d}^4 x \sqrt{-g} \left[ \frac{\Mpl^2}{2}R
+X-V_0  e^{-\lambda \phi/\Mpl}
-\left( e^{Q \phi/\Mpl}-1 \right)\rho_c+\beta Z^2 \right]
+{\cal S}_m\,,
\label{action}
\ee
where $g$ is a determinant of the metric tensor $g_{\mu \nu}$, 
${\cal S}_m$ is the matter action containing the 
contributions of CDM, baryons, and radiation 
with the energy densities $\rho_I$, EOSs $w_I$,  
and squared sound speeds $c_I$, which are 
labeled by $I=c,b,r$ respectively.
We assume that neither baryons nor radiation are 
coupled to the scalar field. 
The action ${\cal S}_m$ of perfect fluids can be 
expressed as a form of the Schutz-Sorkin 
action \cite{Schutz:1977df,Brown:1992kc,DeFelice:2009bx}
\be
{\cal S}_m=-\sum_{I=c,b,r}\int {\rm d}^{4}x \left[
\sqrt{-g}\,\rho_I(n_I)
+ J_I^{\mu} \partial_{\mu} \ell_I  \right]\,,
\ee
where $\rho_I$ depends on the number density $n_I$ 
of each fluid. The current vector field $J_I^{\mu}$ is 
related to $n_I$ as
$n_I=\sqrt{g_{\mu \nu}J_I^{\mu}J_I^{\nu}/g}$, 
with $\ell_I$ being the Lagrange multiplier. 
The fluid four velocity is given by 
\be
u_I^{\mu}=\frac{J_I^{\mu}}{n_I \sqrt{-g}}\,,
\ee
which satisfies the normalization $u_I^{\mu}u_{I\mu}=-1$.
Varying the action (\ref{action}) with respect to $\ell_I$, 
it follows that $\partial_{\mu}J_I^{\mu}=0$. 
In terms of the four velocity, this current conservation 
translates to
\be
u_I^{\mu} \partial_{\mu} \rho_I+\left( \rho_I+P_I 
\right) \nabla_{\mu} u_I^{\mu}=0\,,
\label{continuity}
\ee 
where $P_I=n_I \rho_{I,n}-\rho_I$ is the pressure of 
each fluid.

We discuss the cosmological dynamics on the spatially-flat 
Friedmann-Lema\^itre-Robertson-Walker (FLRW) 
background given by the line element 
\be
{\rm d}s^2=-{\rm d}t^2+a^2(t) 
\delta_{ij} {\rm d}x^i {\rm d} x^j\,,
\ee
where $a(t)$ is the time-dependent scale factor.
On this background we have $u_I^{\mu}=(1,0,0,0)$ and 
$\nabla_{\mu} u_I^{\mu}=3H$, where $H=\dot{a}/a$ is the 
expansion rate of the Universe and a dot denotes 
the derivative with respect to the cosmic time $t$.
{}From Eq.~(\ref{continuity}), we have
\be
\dot{\rho}_I+3H \left( \rho_I+P_I \right)=0\,,
\label{con}
\ee
which holds for each $I=c,b,r$. 
We consider the cosmological dynamics after the CDM 
and baryons started to behave as non-relativistic particles. 
At this epoch, we have $w_c=0$, $w_b=0$, $c_c^2=0$, 
and $c_b^2=0$. The radiation has a usual relativistic 
EOS $w_r=1/3$ with $c_r^2=1/3$.
The gravitational field equations of motion are given by  
\ba
& & 
3M_{\rm pl}^2 H^2=\rho_{\phi}+
e^{Q\phi/\Mpl}\rho_c+\rho_{b}+\rho_r\,,
\label{Eq00}\\
& & 
M_{\rm pl}^2 \left( 2\dot{H}+3H^2 \right)=
-P_{\phi}-\frac{1}{3} \rho_r\,,
\label{Eq11}
\ea
where $\rho_{\phi}$ and $P_{\phi}$ are the scalar-field 
density and pressure defined, respectively, by 
\be
\rho_{\phi}=
\frac{1}{2} q_s \dot{\phi}^2
+V_0  e^{-\lambda \phi/\Mpl}\,,\qquad 
P_{\phi}=\frac{1}{2}q_s \dot{\phi}^2
-V_0  e^{-\lambda \phi/\Mpl}\,,
\ee
with 
\be
q_s \equiv 1+2\beta\,.
\label{qs}
\ee
We require that $q_s>0$ to have a positive 
kinetic term in $\rho_{\phi}$.

The scalar-field equation can be expressed 
in the form 
\be
\dot{\rho}_{\phi}+3H \left( \rho_{\phi}+P_{\phi} 
\right)=-\frac{Q \dot{\phi}}{\Mpl} 
\hat{\rho}_c\,,
\label{phieq}
\ee
where 
\be
\hat{\rho}_c \equiv e^{Q \phi/\Mpl}\rho_c\,.
\label{hrhoc}
\ee
Note that $\hat{\rho}_c$ is the CDM density containing 
the effect of an energy transfer, and the energy flows from CDM to $\phi$ if $\dot{\phi}>0$ with $Q<0$.
{}From Eq.~(\ref{con}), CDM obeys the continuity 
equation $\dot{\rho}_c+3H (\rho_c+P_c)=0$.
In terms of $\hat{\rho}_c$, this equation can be 
expressed as
\be
\dot{\hat{\rho}}_c+3H \hat{\rho}_c=
+\frac{Q \dot{\phi}}{\Mpl} 
\hat{\rho}_c\,.
\label{cdmeq}
\ee
{}From Eqs.~(\ref{phieq}) and (\ref{cdmeq}), it is clear 
that there is the energy transfer between the scalar field 
and CDM, but the momentum exchange between DE and DM 
does not occur at the background level. 
The effect of the coupling $\beta$ appears only 
as the modification to the coefficient of $\dot{\phi}^2$.

To study the background cosmological dynamics, it is 
convenient to introduce the following dimensionless variables 
\be
x_1=\frac{\dot{\phi}}{\sqrt{6}\Mpl H}\,,\qquad 
x_2=\sqrt{\frac{V_0}{3}} \frac{e^{-\lambda \phi/(2\Mpl)}}{\Mpl H}\,,
\ee
and
\be
\Omega_{\phi}=q_s x_1^2+x_2^2\,,\qquad
\Omega_c=\frac{e^{Q\phi/M_{\rm pl}} \rho_c}{3M_{\rm pl}^2 H^2}\,,
\qquad 
\Omega_b=\frac{\rho_b}{3M_{\rm pl}^2 H^2}\,,
\qquad 
\Omega_r=\frac{\rho_r}{3M_{\rm pl}^2 H^2}\,.
\ee
{}From Eq.~(\ref{Eq00}), the density parameters are subject to 
the constraint 
\be
\Omega_c= 1-\Omega_{\phi}-\Omega_b-\Omega_r\,.
\ee
The variables $x_1$, $x_2$, $\Omega_b$, and $\Omega_r$ 
obey the differential equations 
\ba
\frac{\rd x_1}{\rd N} &=& \frac{1}{2} x_1 \left( 6 q_s x_1^2-6 
+3\Omega_c+3\Omega_b+4\Omega_r \right)
+\frac{\sqrt{6}}{2 q_s} \left( \lambda x_2^2
-Q \Omega_c \right)\,,\label{auto1}\\
\frac{\rd x_2}{\rd N}  &=& \frac{1}{2} x_2 \left( 6 q_s x_1^2 
-\sqrt{6} \lambda x_1+3\Omega_c
+3\Omega_b+4\Omega_r \right)\,,\\
\frac{\rd \Omega_b}{\rd N} &=& \Omega_b \left( 6 q_s x_1^2-3 
+3\Omega_c+3\Omega_b+4\Omega_r \right)\,,\\
\frac{\rd \Omega_r}{\rd N} &=& \Omega_r \left( 6 q_s x_1^2-4 
+3\Omega_c+3\Omega_b+4\Omega_r \right)\,,
\label{auto4}
\ea
where $N=\ln a$.
The scalar-field EOS 
$w_{\phi}=P_{\phi}/\rho_{\phi}$ 
and effective EOS 
$w_{\rm eff}=-1-2\dot{H}/(3H^2)$ are 
\be
w_{\phi}=\frac{q_s x_1^2-x_2^2}{q_s x_1^2+x_2^2}\,,\qquad
w_{\rm eff}=-1+2 q_s x_1^2+\Omega_c+\Omega_b
+\frac{4}{3} \Omega_r\,.
\ee

The fixed points with constant values of $x_1$, $x_2$, 
$\Omega_b$, and $\Omega_r$ relevant to the radiation, matter, 
and dark-energy dominated epochs are given, 
respectively, by 
\begin{itemize}
\item Radiation point (A) 
\be
x_1=0\,,\quad 
x_2=0\,,\quad
\Omega_b=0\,,\quad \Omega_r=1\,,\quad 
\Omega_{\phi}=0\,,\quad
w_{\rm eff}=\frac{1}{3}
\label{pointA}\,.
\ee
\item $\phi$MDE point (B) 
\be
x_1=-\frac{\sqrt{6}Q}{3q_s}\,,\quad 
x_2=0\,,\quad
\Omega_b=0\,,\quad \Omega_r=0\,,\quad 
\Omega_{\phi}=w_{\rm eff}=\frac{2Q^2}{3q_s}\,,\quad
w_{\phi}=1\,.
\label{pointB}
\ee
\item Accelerated point (C) 
\be
x_1=\frac{\lambda}{\sqrt{6}q_s}\,,\quad 
x_2=\sqrt{1-\frac{\lambda^2}{6q_s}}\,,\quad
\Omega_b=0\,,\quad \Omega_r=0\,,\quad 
\Omega_{\phi}=1\,,\quad
w_{\phi}=w_{\rm eff}=-1+\frac{\lambda^2}{3q_s}
\label{pointC}\,.
\ee
\end{itemize}
The coupling $Q$ modifies the standard matter era 
through the nonvanishing values of $\Omega_{\phi}$ and $w_{\rm eff}$. 
To avoid the dominance of the scalar-field density over the CDM 
and baryon densities during the $\phi$MDE, we require that  
$\Omega_{\phi} \ll 1$, i.e., 
\be
Q^2 \ll \frac{3}{2} (1+2\beta)\,.
\label{con1}
\ee
To have the epoch of late-time cosmic acceleration 
driven by point (C), we need the condition 
$w_{\rm eff}<-1/3$, i.e., 
\be
\lambda^2<2(1+2\beta)\,.
\ee
Under this condition, we can show that point (C) is stable 
against the homogeneous perturbation if \cite{Amendola:2020ldb}
\be
\lambda (\lambda+Q)<3 (1+2\beta)\,.
\label{con3}
\ee
Provided that the conditions (\ref{con1})-(\ref{con3}) hold, 
the cosmological sequence of fixed points 
(A) $\to$ (B) $\to$ (C) can be realized. 
We refer the reader to Ref.~\cite{Amendola:2020ldb} 
for the numerically integrated background solution.
Taking the limits $Q \to 0$, $\beta \to 0$, and $\lambda \to 0$, 
we recover the background evolution in the $\Lambda$CDM model.

\section{Perturbation equations of motion}
\label{persec}

In Ref.~\cite{Amendola:2020ldb}, the scalar perturbation 
equations of motion were derived without fixing particular gauges.
The perturbed line element containing four scalar 
perturbations $\alpha$, $\chi$, $\zeta$, and $E$ 
on the spatially-flat FLRW background 
is given by 
\be
{\rm d}s^2=-(1+2\alpha) {\rm d}t^2
+2 \partial_i \chi {\rm d}t {\rm d}x^i
+a^2(t) \left[ (1+2\zeta) \delta_{ij}
+2\partial_i \partial_j E \right] {\rm d}x^i {\rm d}x^j\,.
\label{permet}
\ee
Tensor perturbations propagate in the same manner as 
in the $\Lambda$CDM model, 
so we do not consider them in the following.
The scalar field $\phi$ is decomposed into the background 
part $\bar{\phi}(t)$ and the perturbed part $\delta \phi$, as 
\be
\phi=\bar{\phi}(t)+\delta \phi (t,x^i)\,,
\label{phide}
\ee
where we omit the bar from background quantities in the following.

The spatial components of four velocities 
$u_{Ii}=J_{Ii}/(n_I \sqrt{-g})$ 
in perfect fluids are related to the scalar velocity 
potentials $v_I$, as 
\be
u_{Ii}=-\partial_i v_I\,.
\ee
The fluid density is given by 
$\rho_I=\rho_I(t)+\delta \rho_I (t,x^i)$, 
where the perturbed part 
is \cite{Kase:2019veo,Kase:2019mox,Kase:2020hst}
\be
\delta \rho_I=\frac{\rho_{I,n_I}}{a^3} 
\left[ \delta J_I-{\cal N}_I \left( 3\zeta+\partial^2 E 
\right) \right]\,,
\ee
where $\rho_{I,n_I}=\partial \rho_I/\partial n_I$, and 
${\cal N}_I=n_I a^3$ is the background  
particle number of each fluid (which is conserved).

We can construct the following gauge-invariant 
combinations
\ba
&&
\delta \phi_{\rm N}=\delta \phi+\dot{\phi}\left(\chi-a^2 \dot{E}\right)\,,
\qquad 
\delta \rho_{I\rm N}=\delta \rho_I+\dot{\rho}_I \left(\chi-a^2 \dot{E}\right)\,,
\qquad 
v_{I{\rm N}}=v_I+\chi-a^2 \dot{E}\,,\nonumber \\
&&
\Psi=\alpha+\frac{{\rm d}}{{\rm d}t} 
\left( \chi - a^2 \dot{E} \right)\,,\qquad 
\Phi=\zeta+H \left( \chi - a^2 \dot{E} \right)\,.
\label{gava1}
\ea
We also introduce the dimensionless variables 
\be
\delta_{I{\rm N}}= \frac{\delta \rho_{I\rm N}}{\rho_I}\,,\qquad
\delta \varphi_{\rm N}=\frac{H}{\tp}\delta\phi_{\rm N}\,,\qquad
V_{I\rm N}=H v_{I{\rm N}}\,,\qquad 
{\cal K}=\frac{k}{aH}\,,
\label{gava2}
\ee
where $k$ is a comoving wavenumber.
In Fourier space, the linear perturbation equations 
of motion are given by \cite{Amendola:2020ldb}
\ba
& & 
6 q_s x_1^2 \frac{\rd \delta \varphi_{\rm N}}{\rd N}
-6 \frac{\rd \Phi}{\rd N}+6 \left( 
1-q_s x_1^2 \right) \left( \xi \delta \varphi_{\rm N} 
+\Psi \right)-2{\cal K}^2 \Phi
+3 \left( 3\Omega_c+3\Omega_b+4\Omega_r 
\right) \delta \varphi_{\rm N}  \nonumber \\
& &
+3 \left( \Omega_c \delta_{c{\rm N}} 
+\Omega_b \delta_{b{\rm N}}
+\Omega_r \delta_{r{\rm N}}  \right)=0\,,
\label{pereq1}\\
& &
\frac{\rd \Phi}{\rd N}-\Psi-\xi \delta \varphi_{\rm N}
+\frac{3}{2} \left( \Omega_c+4\beta  x_1^2 
\right) \left( V_{c{\rm N}}-\delta \varphi_{\rm N} \right)
+\frac{3}{2} \Omega_b 
\left( V_{b{\rm N}}-\delta \varphi_{\rm N} \right) 
+2\Omega_r 
\left( V_{r{\rm N}}-\delta \varphi_{\rm N} \right)=0\,,\\
& &
\frac{\rd \delta_{I{\rm N}}}{\rd N}
+3 \left( c_I^2-w_I \right) \delta_{I{\rm N}}
+\left( 1+w_I \right) \left( {\cal K}^2 V_{I{\rm N}}
+3 \frac{\rd \Phi}{\rd N} \right)=0\,,\qquad ({\rm for}~I=c,b,r), 
\label{pereq3}\\
& & 
\left( \Omega_c+4 \beta x_1^2 \right) 
\frac{\rd V_{c{\rm N}}}{\rd N}
-\left[ \xi \left( \Omega_c+4 \beta  x_1^2 \right)
-4 \beta x_1^2 (3+2\epsilon_{\phi}) 
-\sqrt{6} Q x_1 \Omega_c
\right] V_{c{\rm N}}
-\Omega_c \Psi \nonumber \\
& &
-4 \beta x_1^2 \frac{\rd \delta \varphi_{\rm N}}{\rd N}
+\left[ 4 \beta x_1 ( \xi -3-2\epsilon_{\phi})
-\sqrt{6}Q \Omega_c \right] x_1
\delta \varphi_{\rm N} =0\,,
\label{pereq4}\\
& &
\frac{\rd V_{I{\rm N}}}{\rd N}
-\left( \xi+3 c_I^2 \right) V_{I{\rm N}} 
-\Psi-\frac{c_I^2}{1+w_I} \delta_{I{\rm N}}=0\,,
\qquad ({\rm for}~I=b,r),
\label{pereq5}\\
& &
\frac{\rd^2 \delta \varphi_{\rm N}}{\rd N^2}
+ \left( 3-\xi +2\epsilon_{\phi} 
\right)\frac{\rd \delta \varphi_{\rm N}}{\rd N}
+\left[ \hat{c}_s^2 {\cal K}^2-\frac{\rd \xi}{\rd N}
-3\xi+\frac{\rd \epsilon_{\phi}}{\rd N}
+\epsilon_{\phi}^2
+(3-\xi)\epsilon_{\phi}
+\frac{3}{q_s} \left( \lambda^2 x_2^2
+Q^2 \Omega_c \right) \right] 
\delta \varphi_{\rm N} \nonumber \\
& &+3\hat{c}_s^2 \frac{\rd \Phi}{\rd N}-\frac{\rd \Psi}{\rd N}
-2\left(3+\epsilon_{\phi}\right)\Psi
-\frac{2\beta}{q_s} \frac{\rd \delta_{c{\rm N}}}{\rd N}
+\frac{\sqrt{6} Q \Omega_c}{2q_s x_1} \delta_{c{\rm N}}=0\,,
\label{delphi} \\
& &
\Psi=-\Phi\,,
\label{pereq6}
\ea
where 
\be
\xi = 
-3q_s x_1^2-\frac{3}{2}\Omega_c-\frac{3}{2}\Omega_b
-2\Omega_r\,,\qquad
\epsilon_{\phi}
= -3+\frac{\sqrt{6}}{2q_s x_1} 
\left( \lambda x_2^2-Q \Omega_{c} \right)\,,\qquad
\hat{c}_s^2=\frac{1}{q_s}\,.
\ee

We can choose any convenient gauges at hand in the 
perturbation Eqs.~(\ref{pereq1})-(\ref{pereq6}). 
For example, the Newtonian gauge corresponds to $\chi=0=E$, 
in which case Eqs.~(\ref{pereq1})-(\ref{pereq6}) can be directly 
solved for the gravitational potentials $\Psi$, $\Phi$ and 
the scalar-field perturbation $\delta \varphi_{\rm N}$.
For the unitary gauge $\delta \phi=0=E$, 
we can introduce the curvature perturbation 
${\cal R}=\Phi-\delta \varphi_{\rm N}$ and the CDM
density perturbation 
$\delta \rho_{c{\rm u}}=
\delta \rho_{c{\rm N}}-\dot{\rho}_c \delta \phi_{\rm N}/\dot{\phi}$ 
as two propagating degrees of freedom. 
These dynamical perturbations have neither ghost nor Laplacian 
instabilities under the following 
conditions \cite{Kase:2019veo,Kase:2019mox,Kase:2020hst}
\ba
q_s &\equiv& 1+2\beta>0\,,
\label{qs} \\
q_c &\equiv& 1+\frac{4\beta x_1^2}{\Omega_c}>0\,,
\label{qc} \\
c_s^2 &\equiv& 
\hat{c}_s^2+\frac{8 \beta^2 x_1^2}
{q_s (4\beta x_1^2+\Omega_c)}>0\,.
\label{cs2}
\ea
Since the CDM effective sound speed vanishes for $c_c^2 \to +0$, 
it does not provide an additional Laplacian 
stability condition. The conditions (\ref{qs})-(\ref{cs2}) 
are independent of the gauge choices.

The evolution of perturbations after the onset of the 
$\phi$MDE can be analytically estimated for
the modes deep inside the sound horizon. 
Under the quasi-static approximation, the dominant 
terms in Eqs.~(\ref{pereq1})-(\ref{pereq6}) are those 
containing ${\cal K}^2$, $\delta_{c{\rm N}}$, 
$\rd \delta_{c{\rm N}}/\rd N$, and $\delta_{b{\rm N}}$.
From Eqs.~(\ref{pereq1}), (\ref{delphi}), and 
(\ref{pereq6}), it follows that 
\be
\Psi=-\Phi \simeq -\frac{3}{2{\cal K}^2} 
\left(  \Omega_c \delta_{c{\rm N}} 
+\Omega_b \delta_{b{\rm N}} \right)\,,\qquad 
\delta \varphi_{\rm N} \simeq \frac{1}{q_s \hat{c}_s^2 {\cal K}^2} 
\left( 2 \beta \frac{\rd \delta_{c{\rm N}}}{\rd N}
-\frac{\sqrt{6} Q \Omega_c}{2x_1}
\delta_{c{\rm N}} \right)\,.
\label{PsiPhi}
\ee
We differentiate Eq.~(\ref{pereq3}) with respect to $N$ 
and then use Eqs.~(\ref{pereq4}) and (\ref{pereq5}) for 
CDM and baryons, respectively. 
On using Eq.~(\ref{PsiPhi}) together with 
the quasi-static approximation, we obtain the 
second-order differential equations of CDM 
and baryons, as \cite{Amendola:2020ldb}
\ba
& &
\frac{\rd^2 \delta_{c{\rm N}}}{\rd N^2}
+\nu \frac{\rd \delta_{c{\rm N}}}{\rd N}
-\frac{3}{2G} \left( G_{cc} \Omega_c \delta_{c{\rm N}}
+G_{cb} \Omega_b \delta_{b{\rm N}} \right) \simeq 0\,,
\label{delceq2} \\
& &
\frac{\rd^2 \delta_{b{\rm N}}}{\rd N^2}+\left( 2+\xi \right) \frac{\rd \delta_{b{\rm N}}}{\rd N}
-\frac{3}{2G} \left( G_{bc} \Omega_c \delta_{c{\rm N}}
+G_{bb} \Omega_b \delta_{b{\rm N}} \right) \simeq 0\,,
\label{delbeq2}
\ea
where 
\be
G_{cc} = \frac{1+r_1}{1+r_2}G\,,\qquad 
G_{cb} = \frac{1}{1+r_2}G\,,\qquad
G_{bc} = G_{bb}=G\,,\label{Gba}
\ee
with
\be
r_1 = \frac{2Q[3Q \Omega_c+2\sqrt{6} \beta x_1
(2+\epsilon_{\phi}+\sqrt{6}Q x_1)]}
{3\Omega_c}\,,\qquad
r_2 = \frac{4\beta (1+2\beta)x_1^2}
{\Omega_c}\,,
\ee
and 
\be
\nu=\frac{4 \beta (1+2\beta)(5+\xi+2\epsilon_{\phi})x_1^2 
+(2+\xi+\sqrt{6}Qx_1)\Omega_c}
{4\beta (1+2\beta) x_1^2+\Omega_c }\,.
\ee
Since $G_{bc}$ and $G_{bb}$ are equivalent to $G$, 
the baryon perturbation is not affected by the 
DE-DM couplings. On the other hand, $G_{cc}$ and $G_{cb}$ 
are different from $G$ for nonvanishing values of 
$Q$ and $\beta$.

During the $\phi$MDE, we obtain 
\be
G_{cc}
=\left( 1+\frac{2Q^2}{1+2\beta} \right)G\,,
\qquad
G_{cb}
= \left[ 1-\frac{8\beta Q^2}{3-2Q^2
+2(3+ 4Q^2)\beta} \right]G\,.
\label{Gcb}
\ee
Under the no-ghost condition (\ref{qs}), we have 
$G_{cc}>G$. So long as the coupling $Q$ is in the range 
$Q^2 \ll 1$, $G_{cb}$ is smaller than $G$.

After the end of the $\phi$MDE, we do not have a 
simple formula for $G_{cc}$. However, assuming that 
$|\beta| \ll 1$ and $|Q| \ll 1$, we find
\be
G_{cc} \simeq  \left( 1+2Q^2-\frac{4\beta x_1^2}{\Omega_c} 
\right)G\,.
\label{Gcc2}
\ee
Since $\Omega_c$ decreases and $x_1^2$ increases at low 
redshifts, the third term in the parenthesis of 
Eq.~(\ref{Gcc2}) dominates over $2Q^2$ to realize 
the value of $G_{cc}$ smaller than $G$. 
Indeed, the numerical simulation in Ref.~\cite{Amendola:2020ldb} 
shows that the growth rate of $\delta_{c{\rm N}}$ 
can be less than the value for $\beta=0$ even 
in the presence of the coupling $Q$.
This suppressed growth of $\delta_{c{\rm N}}$ at low 
redshifts should allow the possibility of reducing 
the $\sigma_8$ tension.

\section{Methodology}
\label{implementation}

We implement our model into the public code \texttt{CAMB}~\cite{Lewis_camb_2000} and simulate 
the evolution of density perturbations with
the background equations to compute the CMB 
and matter power spectra.
In this section, we rewrite the background and perturbation equations of motion in the language of the \texttt{CAMB} code. 
For this purpose, we use the conformal time 
defined by $\tau=\int a^{-1}{\rm d}t$.
The background Eqs.~(\ref{con}), (\ref{Eq00}), 
(\ref{Eq11}), and (\ref{phieq}) can be expressed as
\ba
& & \rho_I'+3\cH \left( \rho_I+P_I \right)=0\,,\qquad 
({\rm for}~~I=c,b,r)\,,\\
& &
3\Mpl^2 \cH^2= \frac12 q_s \phi'^2
+a^2 \left( V_0 e^{-\lambda \phi/\Mpl}
+e^{Q\phi/\Mpl} \rho_c+\rho_b+\rho_r \right)\,,\label{EqBE01}\\
& &
2\Mpl^2 \left( \cH'-\cH^2 \right)=
-q_s \phi'^2-a^2 \left( e^{Q\phi/\Mpl}\rho_c
+\rho_b+\frac{4}{3}\rho_r \right)\,,\label{EqBE02}\\
& & 
q_s \left( \phi''+2 \cH \phi' \right)+\frac{a^2}{\Mpl} 
\left( Q\rho_c e^{Q\phi/\Mpl}
-\lambda V_0 e^{-\lambda \phi/\Mpl} 
\right)=0\,,
\label{EqBE03}
\ea
where a prime represents the derivative with respect 
to $\tau$, and we have introduced the conformal Hubble 
parameter $\cH$ as 
\be
\cH \equiv aH=\dot{a}=\frac{a'}{a}\,.
\ee

For perturbations, we adopt the synchronous 
gauge conditions
\be
\alpha=0\,,\qquad \chi=0\,.
\ee
Following Ma and Bertschinger \cite{Ma:1995ey}, 
we use the notations 
\be
\zeta=-\eta\,,\qquad E=-\frac{h+6\eta}{2k^2}\,,\qquad 
\theta_I=\frac{k^2}{a}v_I\,.
\ee
Then, some of the gauge-invariant variables defined 
in Eqs.~(\ref{gava1}) and (\ref{gava2}) reduce to
\ba
& &
\Psi=\frac{1}{2k^2} \left( h''+\cH h'+6 \eta''+6 \cH \eta' 
\right)\,,\qquad 
\Phi=-\eta+\frac{\cH}{2k^2} \left( h'+6 \eta' \right)\,,\nonumber \\
& &
\delta_{I{\rm N}}=\delta_I-\frac{3\cH}{2k^2} (1+w_I)
(h'+6 \eta')\,,\qquad 
\delta \varphi_{I{\rm N}}=\cH \left( \frac{\delta \phi}{\phi'}
+\frac{h'+6\eta'}{2k^2} \right)\,,\qquad 
V_{I{\rm N}}=\frac{\cH}{k^2} \left( \theta_I +\frac{1}{2}h'
+3 \eta' \right)\,,
\label{corre}
\ea
where $\delta_I \equiv \delta \rho_I/\rho_I$ 
and $w_I \equiv P_I/\rho_I$.  
In the presence of perfect fluids of CDM ($w_c=0=c_c^2$), 
baryons ($w_b=0=c_b^2$), and radiation ($w_r=1/3=c_r^2$), 
we can express the perturbation 
Eqs.~(\ref{pereq1})-(\ref{pereq6}) in the forms
\ba
& &
k^2 \eta-\frac{\cH}{2} h'+\frac{a^2}{2\Mpl^2} 
\left[ \frac{q_s}{a^2}\phi' \delta \phi'
+\frac{1}{\Mpl} \left( Q\rho_c e^{Q\phi/\Mpl}
-\lambda V_0 e^{-\lambda \phi/\Mpl} 
\right) \delta \phi+e^{Q\phi/\Mpl} \rho_c \delta_c
+\rho_b \delta_b+\rho_r \delta_r \right]=0,\\
& &
k^2 \eta'-\frac{a^2}{2\Mpl^2} 
\left[ \frac{k^2}{a^2}\phi' \delta \phi
+\left( \rho_c e^{Q\phi/\Mpl}+\frac{2\beta \phi'^2}{a^2} \right) 
\theta_c
+\rho_b \theta_b+\frac{4}{3}\rho_r \theta_r \right]=0,\\
& &
\delta_c'+\theta_c+\frac{1}{2}h'=0,\\
& &
\delta_b'+\theta_b+\frac{1}{2}h'=0,\\
& &
\delta_r'+\frac{4}{3} \theta_r+\frac{2}{3}h'=0,\\
& &
\theta_c'+\cH \theta_c-\frac{1}{q_s q_c \phi'^2 \Mpl^2} 
\biggl[ q_s (q_c-1)\phi' \Mpl k^2 \delta \phi' 
+\left\{ Q \phi'^2+a^2 (q_c-1)\lambda V_0 e^{-\lambda \phi/\Mpl}
\right\}k^2 \delta \phi \nonumber \\
& &\qquad \qquad \qquad \qquad \qquad~
+\left\{ Q (q_s-2)\phi'^3+3q_s (q_c-1) \cH \phi'^2 \Mpl 
-2a^2(q_c-1)\phi' \lambda V_0 e^{-\lambda \phi/\Mpl} 
\right\}\theta_c \biggr]=0,\\
& &
\theta_b'+\cH \theta_b=0\,,\\
& &
\theta_r'-\frac{k^2}{4}\delta_r=0\,,\\
& &
\delta \phi''+2\cH \delta \phi'
+\frac{k^2 \Mpl^2+a^2 (\lambda^2 
V_0 e^{-\lambda \phi/\Mpl}+Q^2 \rho_c e^{Q\phi/\Mpl})}
{q_s \Mpl^2}\delta \phi
+\frac{\phi'}{2}h'+\frac{2\beta}{q_s}
\phi' \theta_c+\frac{a^2 Q\rho_c e^{Q\phi/\Mpl}}{q_s \Mpl}\delta_c
=0,\label{dphiddot}\\
& &
h''+6 \eta''+2\cH (h'+6\eta')-2\eta k^2=0\,,
\ea
where $q_s$ and $q_c$ are defined by Eqs.~(\ref{qs}) and (\ref{qc}), respectively. The perturbation equations of motion for baryons and radiation are the same as those in $\Lambda$CDM model. 
Thus we modify the equations for CDM and gravitational 
field equations in the \texttt{CAMB} code. 
We also take into account the background and perturbation 
equations of motion for the scalar field, i.e.,  
Eqs.~(\ref{EqBE03}) and (\ref{dphiddot}). 
Note that the CDM velocity is usually set to zero all the time as a result of the gauge fixing condition in \texttt{CAMB} based on the synchronous gauge. 
In the models considered here, CDM has non-zero velocity due to the coupling to $\phi$ in the late Universe. However, we will set $\theta_c=0$ as the initial condition to eliminate 
the gauge degree of freedom, assuming that CDM 
streams freely in the early Universe (i.e., we neglect the interaction between DE and CDM) as in the standard scenario. 

In the background Eqs.~(\ref{EqBE01})-(\ref{EqBE03}), the 
coupling $\beta$ appears through the positive no-ghost 
parameter $q_s=1+2\beta$.
In the limit $q_s \to \infty$, Eq.~(\ref{EqBE03}) shows
that $\phi$ approaches a constant after the onset 
of the $\phi$MDE. This limit corresponds to the 
$\Lambda$CDM model with a constant potential energy. 
Since the parameter space for large values of $q_s$ 
spreads widely, the MCMC chains tend to wander in such 
regions. This actually leads to the loss of information 
about the evolution of the scalar field itself. 
To avoid this, we introduce a set of new variables 
$p_s,\hat\lambda,\hat{Q}$ defined by
\be
p_s \equiv q_s^{-1/2}=\frac{1}{\sqrt{1+2\beta}}\,,\qquad
\hat{\lambda} \equiv p_s\lambda\,,\qquad 
\hat{Q} \equiv p_s Q\,.
\ee
As we discussed in Sec.~\ref{persec}, the growth of matter 
perturbations is suppressed for positive values of $\beta$. 
In the MCMC analysis, we will set the prior
\be
\beta \geq 0\,.
\ee
In this case, the stability conditions 
(\ref{qs})-(\ref{cs2}) are automatically satisfied.
Then, the parameter $p_s$ is in the range $0<p_s \le 1$. 
We will choose the uniform prior for $p_s$. 
In this case, unlike the flat prior for $\beta$, 
the MCMC chain does not wander in the region of 
unphysically large values of $\beta$.

For the parameter $\lambda$, we choose the value 
\be
\lambda>0\,,
\ee
without loss of generality. 
In Eq.~(\ref{EqBE03}), we observe that, for $Q>0$, 
the background scalar field can approach the instantaneous 
minima characterized by the condition 
$Q\rho_c e^{Q\phi/\Mpl}=\lambda V_0 e^{-\lambda \phi/\Mpl}$ 
even during the matter era. Since we would like to study 
the case in which the $\phi$MDE is present, we will 
focus on the coupling range 
\be
Q \le 0\,.
\ee
The same prior was chosen in the MCMC analysis of Refs.~\cite{Pettorino:2013oxa,Planck:2015bue,Gomez-Valent:2020mqn}\footnote{In these papers, 
the sign convention of $Q$ is opposite to ours.}
for the coupled DE-DM model with $Q \neq 0$ and $\beta=0$.

To implement our model in the \texttt{CAMB} code, we 
use the unit $\Mpl=1$ and replace $\phi$ and $\delta\phi$ 
with the following new variables
\be
\phi \equiv p_s \hat{\phi}\,,\qquad 
\delta\phi \equiv p_s \delta{\hat\phi}\,.
\ee
Then, the background scalar-field equation 
can be expressed as 
\be
\hat\phi''+2\mathcal{H}\hat\phi'
+a^2\left(\hat\rho_{c,\hat{\phi}}+V_{,\hat{\phi}} 
\right)=0, \label{effective_evolve_equation}
\ee
where $\hat{\rho}_c=\rho_c e^{\hat{Q}\hat{\phi}}$ 
and $V_{,\hat{\phi}}=\rd V/\rd \hat{\phi}$.
The energy density and pressure of $\hat{\phi}$ read 
$\rho_{\phi}=\hat{\phi}'^2/(2a^2)
+V_0 e^{-\hat{\lambda} \hat{\phi}}$ and 
$P_{\phi}=\hat{\phi}'^2/(2a^2)
-V_0 e^{-\hat{\lambda} \hat{\phi}}$, respectively. 
This means that, at the background level, the effect of the 
momentum transfer can be absorbed into the redefined canonical 
scalar field $\hat{\phi}$. We note that $\hat{\phi}$ mediates 
the energy with CDM through the term $a^2\hat\rho_{c,\hat{\phi}}$ 
in Eq.~(\ref{effective_evolve_equation}).
Using the variables and parameters defined above, the perturbation equations of motion for $\theta_c$ and $\delta\phi$ are now 
expressed as 
\ba
& &
\theta'_c+\mathcal{H}\theta_c-\frac{1-p_s^2}{a^2\hat\rho_cq_c}\left[k^2\hat\phi'\delta\hat\phi'
-a^2k^2\delta\hat\phi V_{,\hat{\phi}}+\left(3\mathcal{H}\hat\phi'
+2a^2 V_{,\hat\phi}\right)\hat\phi'\theta_c\right] -\frac{\hat{Q}}{q_c}\left[k^2p_s^2\delta\hat\phi
+(1-2p_s^2)\hat\phi'\theta_c\right]=0\,,\\
& &
\delta\hat\phi''+2\mathcal{H}\delta\hat\phi'
+\left[p_s^2k^2+a^2\left(V_{,\hat\phi\hat\phi}
+\hat\rho_{c,\hat\phi\hat\phi}\right)\right]\delta\hat\phi
+\left[k\mathcal{Z}+(1-p_s^2)\theta_c\right]\hat\phi'
+a^2\hat\rho_{c,\hat\phi}\,\delta_c=0 \,, 
\ea
where $\mathcal{Z}\equiv h'/(2k)$. 
We will also express the other perturbation equations of motion 
in terms of the new variables introduced above and numerically 
solve them with the background equations. 

\begin{figure}[H]
\centering
\begin{tabular}{cc}
\includegraphics[width=0.5\textwidth]{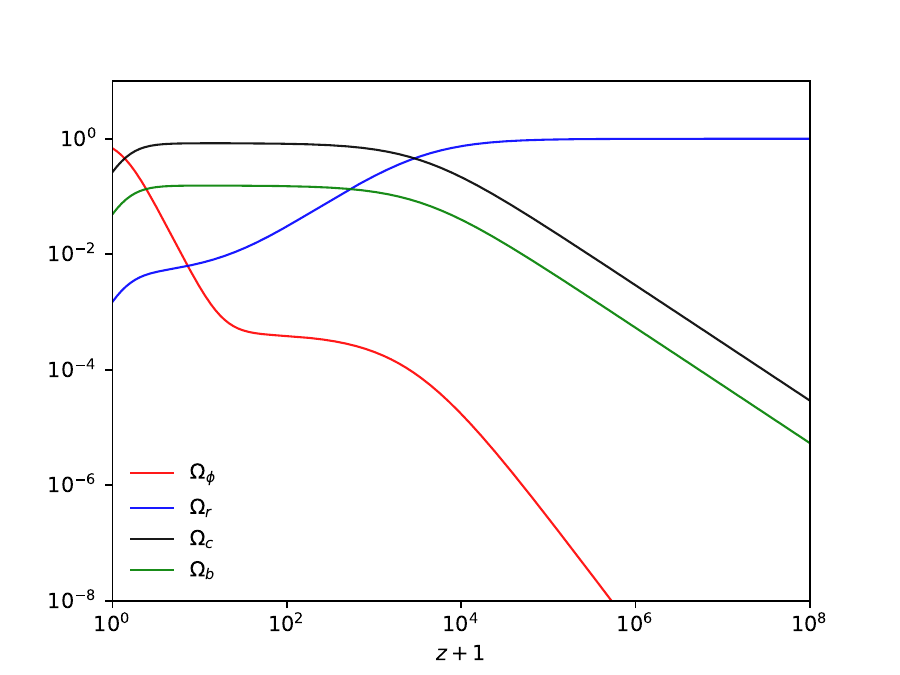}
\includegraphics[width=0.5\textwidth]{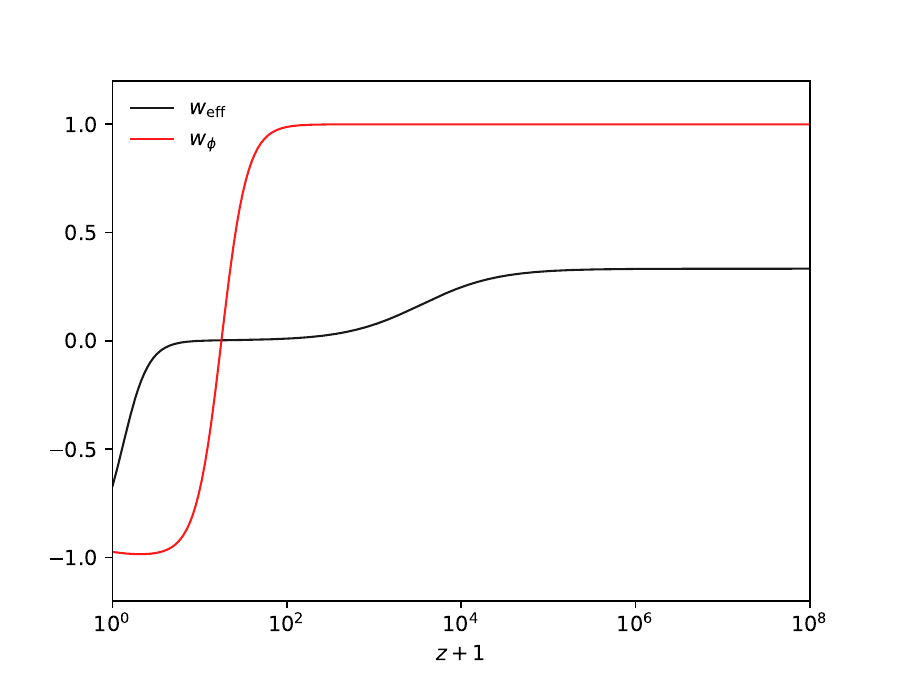}
\end{tabular}
\caption{(Left panel) Evolution of 
$\Omega_\phi$, $\Omega_r$, $\Omega_c$, $\Omega_b$ 
versus $z+1$ ($z$ is the redshift)
for $Q=-0.04$, $\lambda=0.5$, and $\beta=0.4$ 
with today's density parameters 
$\Omega_{c0}=0.25$, $\Omega_{b0}=0.05$, and 
$\Omega_{r0}=1.0 \times 10^{-4}$. 
(Right panel) Evolution of the effective equation of 
state $w_{\rm eff}$ and the scalar-field equation 
of state $w_{\phi}$ for the same model parameters 
and initial conditions as those used in the left.
}
\label{figback}
\end{figure}

\begin{figure}[H]
\centering
\begin{tabular}{cc}
\includegraphics[width=0.5\textwidth]{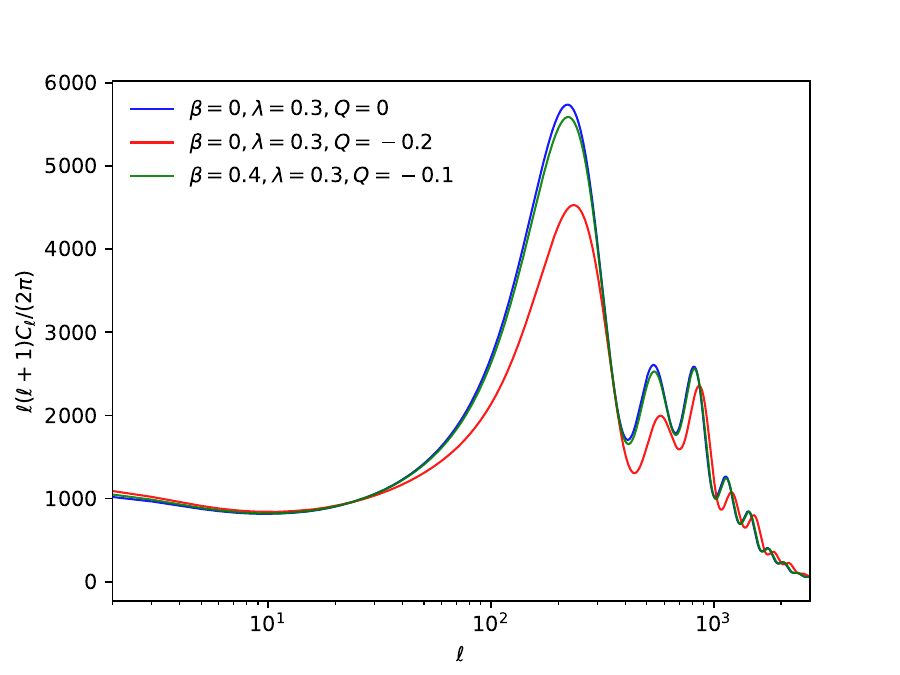}
\includegraphics[width=0.5\textwidth]{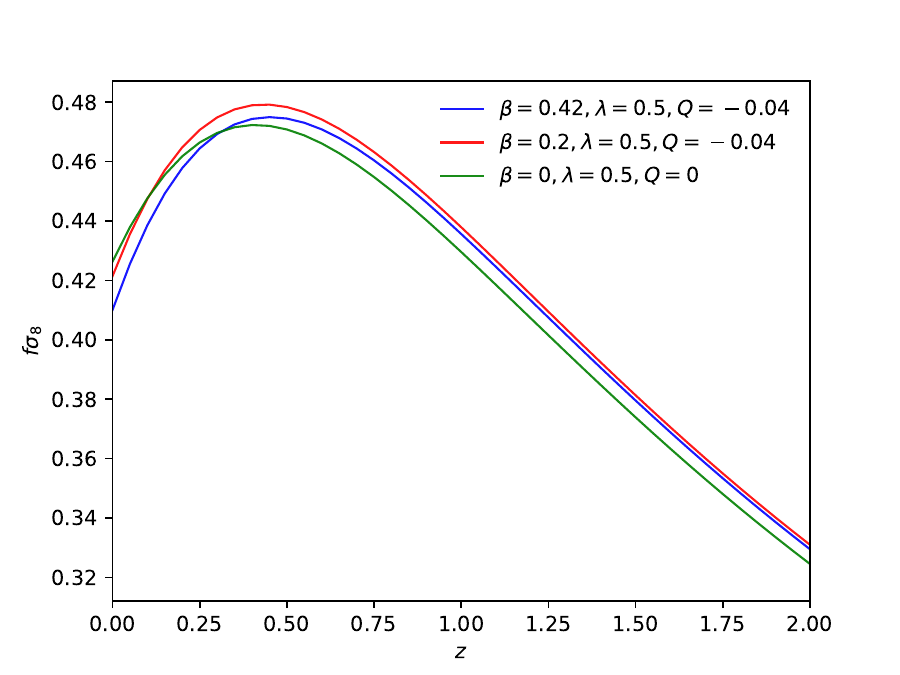}
\end{tabular}
\caption{
(Left panel) Theoretical CMB temperature 
anisotropies versus the multipole $\ell$
for three different model parameter sets: 
(i) $\beta=0$, $\lambda=0.3$, $Q=0$, 
(ii) $\beta=0$, $\lambda=0.3$, $Q=-0.2$, and 
(iii) $\beta=0.4$, $\lambda=0.3$, $Q=-0.1$. 
(Right panel) Evolution of $f\sigma_8$ versus the redshift $z$ for three different cases: 
(i) $\beta=0.42$, $\lambda=0.5$, $Q=-0.04$,
(ii) $\beta=0.2$, $\lambda=0.5$, $Q=-0.04$, 
and (iii) $\beta=0$, $\lambda=0.5$, $Q=0$.
}
\label{figCMBsigma}
\end{figure}

In Fig.~\ref{figback}, we plot the density parameters 
$\Omega_\phi$, $\Omega_r$, $\Omega_c$, $\Omega_b$ (left panel) 
and $w_{\rm eff}$, $w_{\phi}$ (right panel) 
for the model parameters 
$Q=-0.04$, $\lambda=0.5$, and $\beta=0.4$. 
We observe that the solution temporally 
approaches the $\phi$MDE characterized by 
$\Omega_{\phi}=w_{\rm eff}=2Q^2/[3(1+2\beta)]$, 
which is a distinguished feature compared to 
the $\Lambda$CDM model. 
The $\phi$MDE is followed by the epoch of 
cosmic acceleration ($w_{\rm eff}<-1/3$) 
driven by the fixed point (C).

In left panel of Fig.~\ref{figCMBsigma}, we show the CMB 
angular power spectra of temperature anisotropies 
for several different values of $Q$ and $\beta$, 
with $\lambda=0.3$.
Compared to the uncoupled quintessence, there are two main 
effects on CMB induced mostly by the coupling $Q$. 
The first is the shift of acoustic peaks toward larger 
multipoles $\ell$. The multiple $\ell_A$ corresponding to the 
sound horizon $r_{s*}$ at decoupling (redshift $z_*$) 
is given by 
\be
\ell_A=\pi \frac{D_A(z_*)}{r_{s*}}\,,
\label{lA}
\ee
where 
\be
D_A(z_*)=\int_0^{z_*} \frac{1}{H(z)}{\rm d} z
\label{DA}
\ee
is the comoving angular diameter distance, 
and 
\be
r_{s*}=\frac{1}{\sqrt{3}} \int_0^{a_*}
\frac{\rd a}{\sqrt{1+R_s(a)}\,a^2 H(a)}\,,
\label{rs}
\ee
with $R_s(a)=(3\Omega_{b0}/4\Omega_{\gamma 0})a$ 
and $a_*=(1+z_*)^{-1}$ \cite{Hu:1994uz,Amendola:2015ksp}. 
Here, $\Omega_{b0}$ and 
$\Omega_{\gamma 0}$ are today's density parameters of baryons 
and photons, respectively. In our model, there is the $\phi$MDE 
in which the CDM density grows faster toward the higher 
redshift ($\rho_c \propto (1+z)^{3+2Q^2/(1+2\beta)}$) 
in comparison to the uncoupled case ($Q=0$).
Moreover, the scalar-field density $\rho_{\phi}$
scales in the same manner as $\rho_{c}$ during the $\phi$MDE. 
These properties lead to the larger Hubble expansion rate 
before the decoupling epoch, so that the sound horizon 
(\ref{rs}) gets smaller in comparison to the 
uncoupled case. 

The coupling $Q$ can increase the value of $H(z)$ 
from the end of the $\phi$MDE toward the decoupling epoch $z=z_*$, 
which results in the decrease of $D_A(z_*)$.
However, for fixed $H_0$, the increase of $1/r_{s*}$ induced 
by the same coupling typically overwhelms the reduction 
of $D_A(z_*)$ in the estimation of $\ell_A$ in Eq.~(\ref{lA}). 
For the model parameters $Q=0$ with 
$\beta=0$ and $\lambda=0.5$, we obtain the 
numerical values $D_A(z_*)=13.84$~Gpc and 
$r_{s*}=144.40$~Mpc. 
If we change the coupling $Q$ to $-0.2$, 
the two distances change to 
$D_A(z_*)=12.95$~Gpc and 
$r_{s*}=127.20$~Mpc, respectively.
Clearly, the reduction of $r_{s*}$ induced 
by the coupling $Q$ is stronger than the decrease 
of $D_A(z_*)$, which leads to 
the increase of $\ell_A$ from 301.17 
(for $Q=0$) to 319.85 (for $Q=-0.2$).
Hence the larger coupling $|Q|$ leads to the shift of 
CMB acoustic peaks toward smaller scales. 
This effect tends to be significant especially 
for $|Q| \gtrsim 0.1$.
We note that the positive coupling $\beta$ works to 
suppress the factor $2Q^2/(1+2\beta)$ in the 
$(1+z)$-dependent power of $\rho_c$ during the $\phi$MDE.
In comparison to the case $\beta=0$, we need to 
choose larger values of $|Q|$ to have the 
shift of acoustic peaks toward smaller scales.

The second effect of the coupling $Q$ on the CMB temperature 
spectrum is the suppressed amplitude of acoustic peaks. 
The existence of the $\phi$MDE gives rise to the larger 
CDM density $\rho_c$ at decoupling, while the baryon density 
$\rho_b$ is hardly affected. Then, the coupling $Q$ gives rise 
to a smaller ratio $\rho_b/\rho_c$ around $z=z_*$.
For $Q=0$ with $\beta=0$ and $\lambda=0.5$, 
we obtain the numerical value $\rho_b/\rho_c=0.186$, 
while, for $Q=-0.2$ with the same values of $\beta$ 
and $\lambda$, this ratio decreases to 
$\rho_b/\rho_c=0.116$.
This is the main reason for the reduction of the height
of CMB acoustic peaks seen in Fig.~\ref{figCMBsigma}. 
We note that, in the MCMC analysis performed 
in Sec.~\ref{results}, the best-fit value of today's density 
parameter $\Omega_{c0}$ is slightly smaller than the one 
in the $\Lambda$CDM model.
However, for $Q \neq 0$, the increase of $\rho_c$ toward 
the past during the $\phi$MDE results in the larger CDM density 
at decoupling in comparison to the uncoupled case, suppressing the early 
integrated Sachs-Wolfe (ISW) contribution around the first acoustic peak.

From the above discussion, the coupling $|Q|$ of order 
0.1 can reduce the sound horizon at decoupling, whose 
property is required to ease the $H_0$ tension.
However, the same order of $Q$ strongly 
suppresses the amplitude of CMB acoustic peaks. 
To avoid the latter suppression, we require that 
$|Q|$ does not exceed the order of 0.1. 
In Sec.~\ref{results}, we will show that the best-fit 
value of $Q$ constrained from the observational data is 
around $-0.03$. In this case, the sound horizon does not 
decrease significantly in comparison to the case $Q=0$ 
and hence the complete resolution to the $H_0$ tension 
problem is limited.

In the right panel of Fig.~\ref{figCMBsigma}, we show the evolution 
of $f \sigma_8$ for several different model parameters, where $f=\dot{\delta}_m/(H \delta_m)$ is the growth rate of matter density 
contrast (incorporating both CDM and baryons) and $\sigma_8$ is the amplitude of matter over-density at the comoving $8 h^{-1}$~Mpc scale 
($h$ is the normalized Hubble constant $H_0 = 100\,h$ km/s/Mpc).
We find that the large coupling $\beta$ induces the suppression 
for the growth rate of matter perturbations at low redshifts. 
This is also the case even in the presence of 
the coupling $Q$ of order $-0.01$. 
This result is consistent with the analytic estimation 
for the growth of perturbations discussed in Sec.~\ref{persec}.

\section{Results and discussion}
\label{results}

We are now going to place observational constraints on our model 
by using the MCMC likelihood \texttt{CosmoMC}~\cite{Lewis_cosmomc_2014}. 
In our analysis, we will exploit the 
following data sets.

\noindent (i) The CMB data containing TT, TE, EE+lowE from Planck 2018~\cite{planck_2018}, and the full-shape large-scale 
structure data from the 12-th data release of 
SDSS~\cite{sdss_dr12}. We also run the MCMC code by 
using the 16-th SDSS data, but we did not find any obvious 
difference from the constraints derived below. 

\noindent (ii) The Phantheon supernovae samples containing 
$1048$ type Ia supernovae magnitudes with redshift in the range of $0.01<z<2.3$~\cite{pantheon_2018}, which are commonly used 
to constrain the property of late-time cosmic acceleration.

\noindent (iii) The 1-st year DES results~\cite{DES_1yr}, which 
are the combined analyses of galaxy clustering and weak gravitational lensing.

We stop the calculations when the Gelman-Rubin statistic $R-1 \sim 0.01$ is reached.
In Fig.~\ref{fig_MCMCTriangle} and Table~\ref{MCMCResultsTable},  
we present the results of observational constraints on 
our model parameters.

\if\baoonly1
\begin{figure}[H]
\centering
\begin{tabular}{c}
\includegraphics[width=\textwidth]{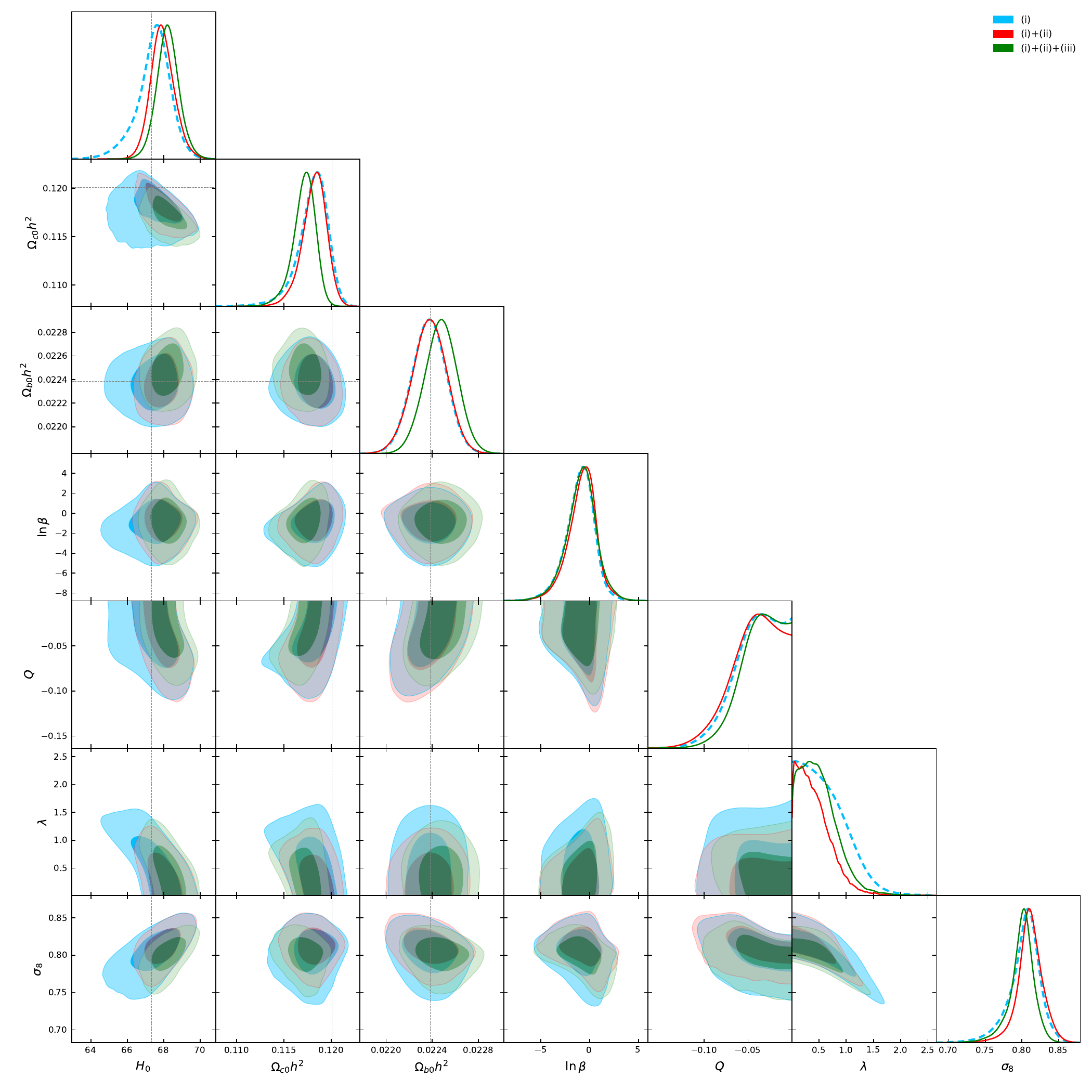}
\end{tabular}
\caption{Triangle plot for the 1-dimensional 
marginalized distributions on individual parameters 
and the 1$\sigma$ and 2$\sigma$ 2-dimensional contours. 
The blue dashed lines represent constraints by 
the Planck 2018~\cite{planck_collaboration_planck_2020} 
and 12-th SDSS data sets, which we call (i).
The red and green solid lines correspond to constraints 
when the data sets (ii) and (ii)+(iii) are combined with (i), 
respectively.
}
\label{fig_MCMCTriangle}
\end{figure}
\else
\begin{figure}[H]
\centering
\begin{tabular}{c}
\includegraphics[width=\textwidth]{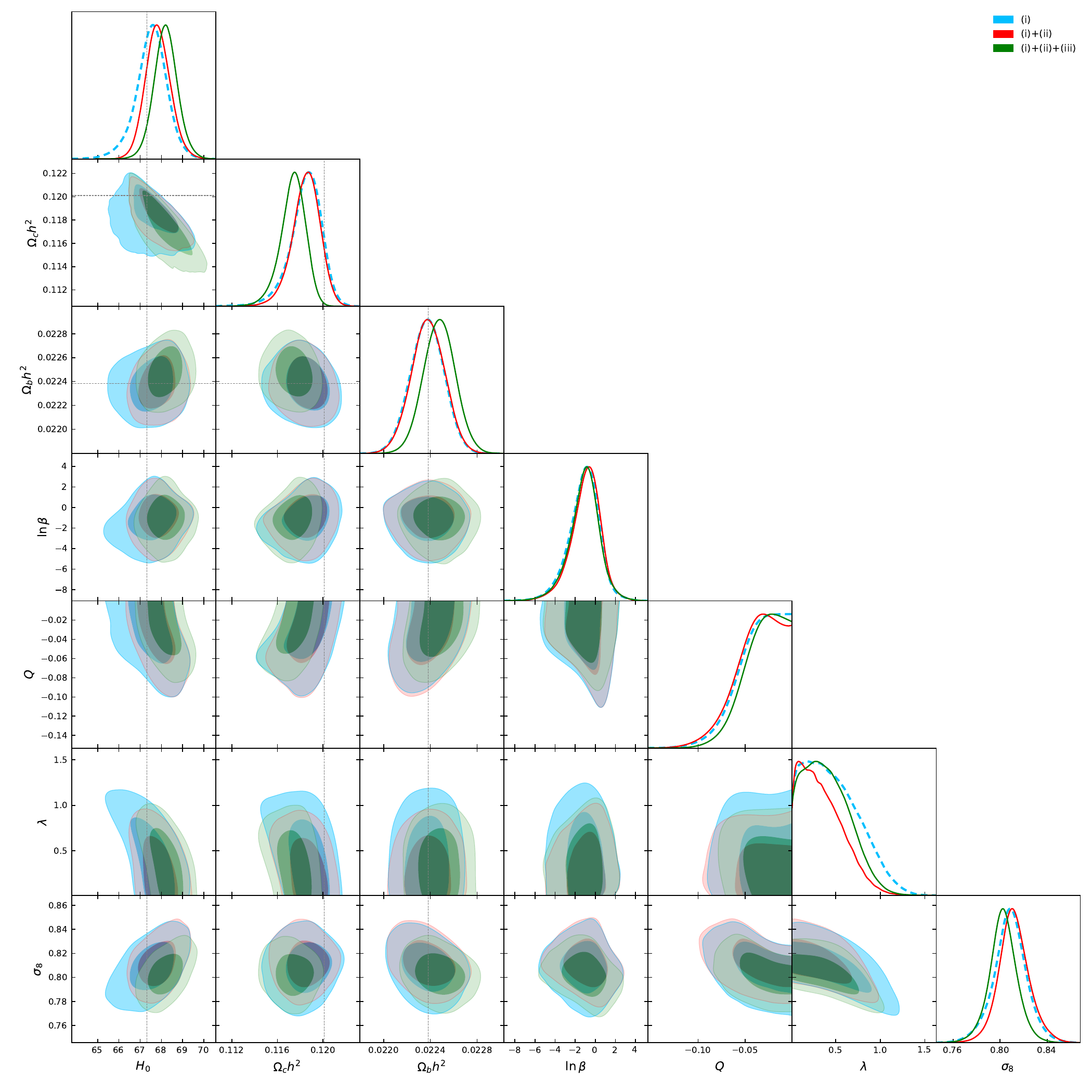}
\end{tabular}
\caption{Triangle plot for the 1-dimensional 
marginalized distributions on individual parameters 
and the 1$\sigma$ and 2$\sigma$ 2-dimensional contours. 
The blue dashed lines represent constraints by 
the Planck 2018~\cite{planck_collaboration_planck_2020} 
and 12-th SDSS data sets, which we call (i).
The red and green solid lines correspond to constraints 
when the data sets (ii) and (ii)+(iii) are combined with (i), 
respectively.
}
\label{fig_MCMCTriangle}
\end{figure}
\fi

\begin{table*}[t]
\caption{Priors, mean values, best-fit values and 1$\sigma$ 
errors of the model parameters $\ln \beta$, 
$\lambda$, $Q$ and cosmological parameters 
$H_0$, $\Omega_{c0} h^2,\Omega_{b0} h^2$, $\sigma_8$, where 
$\Omega_{c0}$ and $\Omega_{b0}$ are today's density parameters 
of CDM and baryons respectively. 
The third, fourth, and fifth columns correspond to the constraints 
derived by the data sets (i), (i)$+$(ii), and (i)$+$(ii)$+$(iii), 
respectively.
}
\begin{center}
\begin{ruledtabular}
\begin{tabular}{ccccc}
Parameters & Priors & mean (best fit) (i) & mean (best fit) (i)+(ii) & mean (best fit) (i)+(ii)+(iii)\\
\hline
$H_0$~[km/s/Mpc] & $[20,100]$ & $67.52\,(67.17)^{+0.76}_{-0.64}$ & $67.84\,(67.87)^{+0.56}_{-0.61}$ & $68.20\,(67.84)^{+0.54}_{-0.55}$ \\

$\Omega_{c0}h^2$ & $[0.001, 0.99]$ & $0.11848\,(0.11861)^{+0.00155}_{-0.00105}$ & $0.11850\,(0.11904)^{+0.00132}_{-0.00106}$ & $0.11735\,(0.11882)^{+0.00122}_{-0.00092}$ \\

$\Omega_{b0}h^2$ & $[0.005, 0.1]$ & $0.02237\,(0.02240)^{+0.00015}_{-0.00014}$ & $0.02238\,(0.02235)^{+0.00015}_{-0.00014}$ & $0.02248\,(0.02241)^{+0.00014}_{-0.00014}$ \\

$\ln\beta$ & * & $-1.1546\,(-2.7365)^{+1.6471}_{-1.2185}$ & $-0.9538\,(1.0231)^{+1.5897}_{-1.1714}$ & $-1.1012\,(0.4111)^{+1.5576}_{-1.2456}$ \\

$\lambda$ & $[0.1,\infty]$ & $0.4650\,(0.5043)^{+0.1421}_{-0.4459}$ & $0.3517\,(0.2875)^{+0.0987}_{-0.3391}$ & $0.4042\,(0.3376)^{+0.1370}_{-0.3673}$ \\

$Q$ & $[-\infty,0]$ & $-0.0347\,(-0.0209)^{+0.0347}_{-0.0093}$ & $-0.0366\,(-0.0716)^{+0.0366}_{-0.0096}$ & $-0.0312\,(-0.0036)^{+0.0312}_{-0.0085}$ \\

$\sigma_8$ & * & $0.8084\,(0.8065)^{+0.0140}_{-0.0130}$ & $0.8118\,(0.8101)^{+0.0118}_{-0.0132}$ & $0.8026\,(0.8002)^{+0.0114}_{-0.0112}$
 \label{MCMCResultsTable}
 \end{tabular}
 \end{ruledtabular}
 \end{center}
\end{table*}

First, let us discuss constraints on the parameter $\beta$. 
In Table \ref{MCMCResultsTable}, the bounds on $\beta$ (68\,\% CL) 
constrained by different data sets are presented 
in terms of the log prior. 
From the joint analysis based on the data sets 
(i)$+$(ii)$+$(iii), this bound translates to 
\be
\beta=0.332^{+1.246}_{-0.237} 
\qquad (68\,\%\,{\rm CL})\,,
\ee
where $0.332$ is the mean value. 
Since $\beta$ is constrained to be larger 
than 0.095 at 1$\sigma$, there is an interesting observational signature of the momentum exchange between DE and DM. 
Even with the analysis of the data set (i) or 
with the data sets (i)$+$(ii), the $1\sigma$ lower limits on $\beta$ are close to the value 0.1. 
In other words, adding the DES data 
to the data sets (i)$+$(ii) hardly modifies the constraint on $\beta$.
Hence the Planck CMB data combined with the SDSS data already show the signature of the momentum transfer. 
We note that this result is consistent with the 
likelihood analysis of Refs.~\cite{Pourtsidou:2016ico,Linton:2017ged,Chamings:2019kcl,Linton:2021cgd} performed for the model $Q=0$, where the joint analysis 
based on the CMB and galaxy clustering data favour nonvanishing 
values of $\beta$.

With the data sets (i)$+$(ii)$+$(iii), 
we also obtain the following 2$\sigma$ bound  
\ba
0.011<\beta< 7.526 \qquad (95\,\%\,{\rm CL})\,.
\ea
Since the lower limit of $\beta$ is as small as $0.01$, 
this value is not significantly distinguished from $\beta=0$. 
This means that the evidence for the momentum transfer can be confirmed at 68\,\% CL, but not firmly at 
95\,\% CL, with the current observational data.
We note that the mean value of $\sigma_8$ constrained by the data sets (i)$+$(ii)$+$(iii) is $0.8026$, which is smaller than the Planck 2018 bound $\sigma_8=0.8111 \pm  0.0060$ \cite{Planck:2018vyg} 
derived for the $\Lambda$CDM model. 
Thus, in our model, the $\sigma_8$ tension between the CMB and other measurements is alleviated by the momentum transfer. 
This property is mostly attributed to the fact that the 
growth rate of $\delta_c$ at low redshifts is suppressed 
by the positive coupling $\beta$.

The other coupling constant $Q$, which mediates the energy transfer 
between DE and DM, is constrained to be 
\be
Q =-0.0312^{+0.0312}_{-0.0085} \qquad (68\,\%\,{\rm CL})\,,
\label{Qcon}
\ee
where $-0.0312$ is the mean value. 
With the data sets (i) alone, the 1-dimensional probability distribution of $Q$ does not have a particular peak at $Q<0$. 
As we see in Fig.~\ref{fig_MCMCTriangle}, 
the MCMC analysis based on the 
full data sets gives rise to a peak in the 
1-dimensional distribution of $Q$ 
around $-0.03$.
Since the vanishing coupling ($Q=0$) is within 
the $1\sigma$ contour, we do not have strong 
observational evidence that the nonvanishing 
value of $Q$ is favored over the $Q=0$ case.
However, it is interesting to note that 
the current data give rise to the probability 
distribution of $Q$ with a peak at $Q<0$.

In Refs.~\cite{Pettorino:2013oxa,Planck:2015bue,Gomez-Valent:2020mqn}, 
the couplings $|Q|$ slightly smaller than 
the mean value of (\ref{Qcon}) were obtained by 
the MCMC analysis with several data sets 
for the coupled dark energy model with $\beta=0$. 
In our model, we have 
$\Omega_{\phi}=w_{\rm eff}=2Q^2/[3(1+2\beta)]$ during the 
$\phi$MDE, so both $\Omega_{\phi}$ and $w_{\rm eff}$ 
are suppressed by the positive coupling $\beta$. 
This allows the larger values of $|Q|$ in comparison to 
the case $\beta=0$. Still, the coupling $|Q|$ exceeding the 
order $0.1$ is forbidden from the data because of the 
significant changes of heights and positions 
in CMB acoustic peaks (see Fig.~\ref{figCMBsigma}).

The parameter $\lambda$ is related to the slope of 
the scalar-field potential. 
To realize the DE equation of state closer to $-1$ at late times,
we require that $\lambda$ can not be significantly 
away from 0. From the MCMC analysis with 
the data sets (i)$+$(ii)$+$(iii), 
we obtain the upper limit 
\be
\lambda<0.5412  \qquad (68\,\%\,{\rm CL})\,.
\ee
The upper limit on $\lambda$ is mostly 
determined by the Planck and SN Ia data sets.
We also remark that, for larger $\lambda$, the distance to the CMB last scattering surface is reduced. To compensate this property, we require smaller values of $H_0$. 
This explains the tendency for blue contours seen in the $\lambda$-$H_0$ plane. Thus, the smaller values of  $\lambda$ are favored from the viewpoint of increasing $H_0$.

In Fig.~\ref{fig_MCMCTriangle}, we find that today's CDM density parameter $\Omega_{c0}$ is constrained to be smaller than the Planck 2018 bound $\Omega_{c0}h^2=0.120 \pm 0.001$ 
derived for the $\Lambda$CDM model \cite{Planck:2018vyg}.
In spite of this decrease of $\Omega_{c0}$, 
the CDM density evolves as $\rho_c \propto (1+z)^{3+2Q^2/(1+2\beta)}$ during the 
$\phi$MDE and hence $\Omega_{c}$ at decoupling can be increased 
by the nonvanishing coupling $Q$.
We note that today's baryon density parameter 
$\Omega_{b0}$ is only slightly larger than the 
Planck 2018 bound $\Omega_{b0}=0.0224 \pm 0.0001$ 
(see Fig.~\ref{fig_MCMCTriangle}). 
Then, the nonvanishing coupling $Q$ hardly modifies the value of $\Omega_{b}$ at $z=z_*$ in 
comparison to the case $Q=0$.
Since the ratio $\Omega_{b}/\Omega_{c}$ at decoupling is decreased by the coupling $|Q|$ larger than the order $0.01$, this suppresses the height of CMB acoustic peaks. 
The MCMC analysis with the CMB data alone already 
places the bound $|Q|<0.1$ at 95\,\%\,CL.

As we discussed in Sec.~\ref{implementation}, 
the nonvanishing coupling $Q$ reduces the sound horizon $r_{s*}$ at $z=z_*$. 
This leads to the shift of CMB acoustic peaks 
toward smaller scales.
To keep the position of the multipole $\ell_A$ 
corresponding to the sound horizon, 
we require that the comoving angular diameter 
distance $D_A(z_*)$ appearing in the numerator 
of Eq.~(\ref{lA}) should be also reduced.
We can express Eq.~(\ref{DA}) as 
$D_A(z_*)=H_0^{-1} \int_0^{z_*}E^{-1}(z) {\rm d}z$, 
where $E(z)=H(z)/H_0$. 
In the $\Lambda$CDM model we have 
$E(z)=[\Omega_{m0} (1+z)^3+\Omega_{\Lambda}+\Omega_{r0} 
(1+z)^4]^{1/2}$, where $\Omega_{m0}=\Omega_{c0}+\Omega_{b0}$.
In our model, the CDM density parameter during 
the $\phi$MDE has the dependence 
$\Omega_{c0} (1+z)^{3+2Q^2/(1+2\beta)}$ instead of 
$\Omega_{c0} (1+z)^{3}$, 
together with the scaling behavior of 
$\rho_{\phi}$ with $\rho_{c}$.
Then, the coupling $Q$ leads to the increase of $E(z)$ 
from the end of $\phi$MDE to the decoupling epoch, so 
that the integral $\int_0^{z_*}E^{-1}(z) {\rm d}z$ 
is decreased. This property is different from the early 
DE scenario of Ref.~\cite{Poulin:2018cxd}, where the 
energy density of early DE quickly decays after the 
recombination epoch.

\begin{figure}[H]
\centering
\begin{tabular}{cc}
\includegraphics[width=0.55\textwidth]{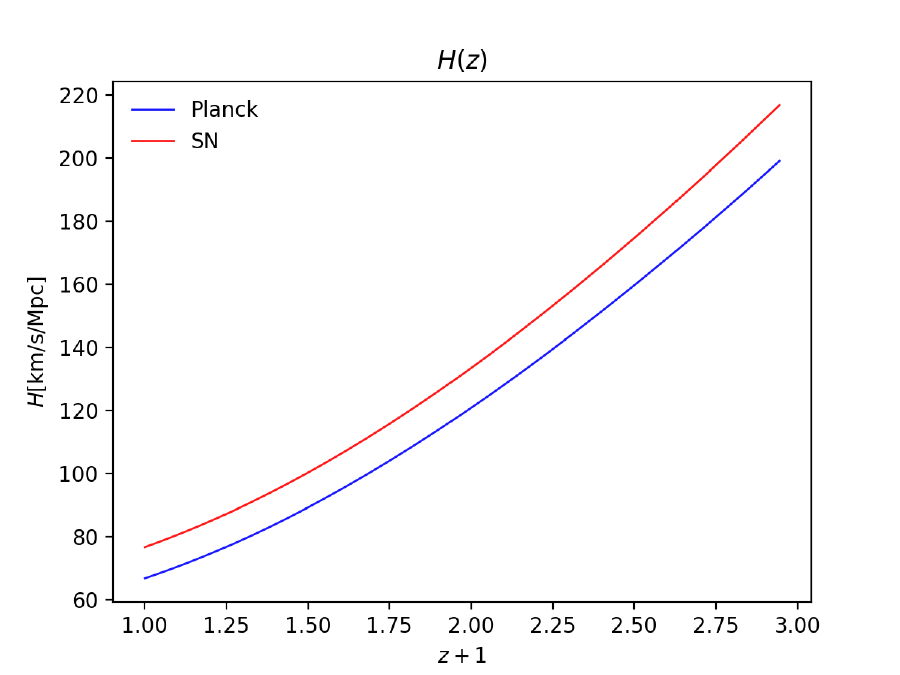}
\end{tabular}
\caption{Evolution of the Hubble expansion rate $H(z)$ 
at low redshifts for the best-fit model parameters 
constrained from the Planck 2018 data alone (blue) 
and from the SN Ia data alone (red). 
}
\label{figHz}
\end{figure}

In our model, increasing the value of $H_0$ also reduces 
$D_A(z_*)$, so it can compensate the reduction of $r_{s*}$. 
However, the integral $\int_0^{z_*}E^{-1}(z) {\rm d}z$ is 
already decreased at some extent by the existence of 
the $\phi$MDE. In this sense, there is the limitation 
for realizing $H_0$ significantly larger than the value 
obtained for $Q=0$.
The observational constraint on $H_0$ derived 
by the data set (i) for the model with $Q=0$ is 
consistent with the Planck 2018 bound 
$H_0=67.27 \pm 0.60$~km/s/Mpc. 
In the presence of the negative coupling $Q$, 
the likelihood region in the $Q$-$H_0$ plane 
shown in Fig.~\ref{fig_MCMCTriangle} shifts toward 
larger values of $H_0$. 
This is particularly the case for adding 
the data sets (ii) and (iii) to the Planck data. 
With the full data sets (i)$+$(ii)$+$(iii), 
the Hubble constant is constrained to be 
\be
H_0=68.20^{+0.54}_{-0.55}~{\rm km/s/Mpc} 
\qquad (68\,\%\,{\rm CL})\,,
\label{H0bo}
\ee
whose mean value is larger than the one derived 
for the $\Lambda$CDM model with  
the Planck 2018 data alone.
However, it is not possible to reach the region 
$H_0>70$~km/s/Mpc due to the limitation of reducing
$D_A(z_*)$ by increasing the value of $H_0$. 
We also carried out the MCMC analysis for the $\Lambda$CDM model and obtained the bound $H_0=68.16^{+0.37}_{-0.38}~{\rm km/s/Mpc}$ 
with the full data sets (i)$+$(ii)$+$(iii).
The $1\sigma$ upper limit of the constraint (\ref{H0bo}) is 
only slightly larger than that of the $\Lambda$CDM bound. 
In Fig.~\ref{figHz}, we plot the evolution of the Hubble expansion rate versus $z+1$ for 
the best-fit model parameters 
constrained from the Planck 2018 data 
alone and from the SN Ia data alone. 
Clearly, there are discrepancies for 
the low-redshift values of $H(z)$ 
obtained from the CMB and SN Ia data. 
This is related to the fact that 
the coupling $|Q|$ of order 0.01 is not 
sufficient to reduce the sound horizon at decoupling significantly for the purpose of obtaining large values of $H_0$.
Hence the Hubble tension problem between the Planck 2018 data 
and those constrained by the direct measurements of $H_0$ still 
persists in our coupled DE scenario.

Albeit the difficulty of resolving the Hubble tension problem, 
the fact that the probability distribution of $Q$ has  
a peak around $-0.03$ is an interesting property 
of our model. Moreover, there are observational signatures 
of the momentum transfer with $\beta>0$ between DE and DM at 68\,\% CL. 
The coupling $\beta$ can alleviate the $\sigma_8$ tension without 
spoiling the existence of the $\phi$MDE.  

\section{Conclusions}
\label{consec}

In this paper, we put observational constraints on an interacting 
model of DE and DM given by the action (\ref{action}). 
Since our model has a concrete Lagrangian, the background and 
perturbation equations of motion are unambiguously fixed 
by the variational principle.
This is not the case for many coupled DE-DM models 
studied in the literature, in which the interacting terms 
are added to the background equations by hands.
In our model, the DE scalar field $\phi$ and the CDM 
fluid mediate both energy and momentum transfers, 
whose coupling strengths are characterized by the 
constants $Q$ and $\beta$, respectively. 
We considered an exponential potential 
$V(\phi)=V_0 e^{-\lambda \phi/\Mpl}$ of the scalar field 
to derive late-time cosmic acceleration, 
but the different choice of quintessence potentials should not  
affect the observational constraints on $Q$ 
and $\beta$ significantly. 

The coupling $Q$ can give rise to the $\phi$MDE during which 
the scalar-field density parameter $\Omega_{\phi}$ and 
the effective equation of state $w_{\rm eff}$ are 
nonvanishing constants, such that 
$\Omega_{\phi}=w_{\rm eff}=2Q^2/[3(1+2\beta)]$.
In this epoch, the CDM density grows as 
$\rho_c \propto (1+z)^{3+2Q^2/(1+2\beta)}$ 
toward the past and hence the value of $\rho_c$ 
at CMB decoupling can be increased by the coupling $Q$. 
Since this enhances the Hubble expansion rate 
in the past, the sound horizon $r_{s*}$ at decoupling 
(redshift $z_*$) gets smaller. Moreover, the ratio 
between the baryon and CDM densities, $\rho_b/\rho_c$, 
is suppressed at $z=z_*$ due to the increase of 
$\rho_c$ induced by the presence of the $\phi$MDE. 
These modifications shift the positions and heights of 
acoustic peaks of CMB temperature anisotropies, 
so that the coupling $Q$ can be tightly
constrained from the CMB data.

The effect of momentum transfers on the dynamics of 
perturbations mostly manifests itself for the evolution  
of CDM density contrast $\delta_c$ at low redshifts. 
For $\beta>0$, the growth of $\delta_c$ is suppressed 
due to the decrease of an effective gravitational coupling 
$G_{\rm eff}$ on scales relevant to the galaxy clustering.
The coupling $Q$ enhances the value of $G_{\rm eff}$ 
through the energy transfer between DE and DM. 
However, the reduction of $G_{\rm eff}$ induced 
by positive $\beta$ typically overwhelms the increase 
of $G_{\rm eff}$ for the redshift $z \lesssim 1$. 
Hence the growth rate of CDM perturbations is suppressed 
in comparison to the $\Lambda$CDM model.

We carried out the MCMC analysis for our model by using the 
observational data of Planck 2018 \cite{planck_2018}, 
12-th SDSS, Phantheon supernovae samples, and 1-year DES.
The coupling $\beta$ is constrained to be in the range 
$\beta=0.332^{+1.246}_{-0.237}$ (68\,\%\,CL) by using all 
the data sets. Since the $\beta=0$ case is outside the 
$1\sigma$ observational contour, there is an interesting 
observational signature of the momentum transfer between 
DE and DM. This is an outcome of the suppressed growth 
of $\delta_c$ at low redshifts, thereby easing the 
$\sigma_8$ tension. Indeed, we found that the mean value of 
$\sigma_8$ constrained by the full data is $0.8026$, 
which is smaller than the best-fit value $0.8111$ derived for the 
$\Lambda$CDM model with the Planck data alone.

For the coupling characterizing the energy transfer, we obtained 
the bound $Q=-0.0312^{+0.0312}_{-0.0085}$ (68\,\%\,CL) by the 
analysis with full data sets. While the $Q=0$ case is within the 
$1\sigma$ observational contour, there is a peak for the 
probability distribution of the coupling at a negative 
value of $Q$. This result is consistent with the likelihood 
analysis performed for the model with 
$\beta=0$ \cite{Pettorino:2013oxa,Planck:2015bue,Gomez-Valent:2020mqn}, 
but now the constrained values of $|Q|$ get larger.  
This increase of $|Q|$ is mostly attributed to the fact that 
the effective equation of state during the $\phi$MDE 
is modified to $w_{\rm eff}=2Q^2/[3(1+2\beta)]$ through the 
coupling $\beta$. In comparison to the momentum transfer, 
we have not yet detected significant observational signatures 
of the energy transfer, but the future high-precision data 
will clarify this issue.

The presence of the coupling $Q$ reduces the sound horizon $r_{s*}$ 
at decoupling, thereby increasing the multipole $\ell_A$ 
defined in Eq.~(\ref{lA}). To keep the position of CMB acoustic 
peaks, we require that the comoving angular diameter distance 
$D_A(z_*)$ from $z=0$ to $z=z_*$ decreases. 
During the $\phi$MDE, the Hubble expansion rate increases 
due to the enhancement of $\rho_c$ induced by the energy transfer.
Since this leads to the decrease of $D_A(z_*)$, the further reduction 
of $D_A(z_*)$ by the choice of larger values of $H_0$ is quite limited 
in our model. From the MCMC analysis of full data sets 
we obtained the bound $H_0=68.20^{+0.54}_{-0.55}~{\rm km/s/Mpc}$, 
whose mean value is larger than the one derived for the 
$\Lambda$CDM model with the Planck 2018 data alone. 
However, the Hubble constant $H_0$ does not exceed the value 
$70~{\rm km/s/Mpc}$, so the Hubble tension problem 
is not completely resolved in our scenario.

It is still encouraging that the current data support 
signatures of the interaction between DE and DM. 
We expect that upcoming observational data like those from 
the Euclid satellite will place further tight constraints 
on the couplings $\beta$ and $Q$. 
Along with the $H_0$ tension problem, we hope that we 
will be able to approach the origins of DE and DM and 
their possible interactions in the foreseeable future.

\section*{Acknowledgements}

XL is supported by the National Natural Science Foundation of China under Grants Nos.~11920101003, 12021003 and 11633001, 
and the Strategic Priority Research Program of the Chinese 
Academy of Sciences, Grant No.~XDB23000000. 
ST is supported by the Grant-in-Aid for Scientific Research 
Fund of the JSPS No.~22K03642 and Waseda University 
Special Research Project No.~2023C-473. 
KI is supported by the JSPS grant number 21H04467, JST FOREST Program JPMJFR20352935, and by JSPS Core-to-Core Program 
(grant number:JPJSCCA20200002, JPJSCCA20200003). 

\bibliographystyle{mybibstyle}
\bibliography{bib}

\end{document}